\documentclass[
aps,10pt,
prl,
preprintnumbers,amsmath,amssymb,
reprint,
nofootinbib,
a4paper,
twocolumn
]{revtex4-2} 

\usepackage[utf8]{inputenc}
\usepackage[T1]{fontenc}
\usepackage{lmodern}
\usepackage{amsmath,amssymb,amsfonts}
\usepackage{graphicx}
\usepackage{hyperref}
\usepackage{xcolor}
\usepackage{mathtools}
\usepackage{bm}
\usepackage{orcidlink} 

\hypersetup{
	colorlinks=true,
	linkcolor=blue,
	citecolor=blue,
	urlcolor=blue
}

\newcommand\nn{\nonumber\\}

\newcommand\bb{\bibitem}
\newcommand{\bma}{\left(\begin{array}}
	\newcommand{\ema}{\end{array}\right)}
\newcommand{\be}{\begin{equation}}
	\newcommand{\ee}{\end{equation}}
\newcommand{\ben}{\begin{equation*}}
	\newcommand{\een}{\end{equation*}}
\newcommand{\ba}{\begin{eqnarray}}
	\newcommand{\ea}{\end{eqnarray}}
\newcommand{\ban}{\begin{eqnarray*}}
	\newcommand{\ean}{\end{eqnarray*}}
\newcommand{\bs}{\begin{subequations}}
	\newcommand{\es}{\end{subequations}}
\newcommand{\bc}{\begin{center}}
	\newcommand{\ec}{\end{center}}

\def\ds{d_{\rm S}}
\def\dh{d_{\rm H}}
\def\dw{d_{\rm W}}

\newcommand{\Pl}{{\text{\tiny Pl}}}

\newcommand{\Mpl}{M_\Pl}

\newcommand{\au}[2]{#1.~#2}
\newcommand{\arX}[1]{\href{http://arxiv.org/abs/#1}{{\cob arXiv:#1}}}
\newcommand{\oarX}[1]{\href{http://arxiv.org/abs/#1}{{\cob #1}}}

\newcommand{\book}[5]{\emph{#1}, #2, #3, #4 (#5)}
\newcommand{\doin}[6]{\href{https://doi.org/#1}{{\cob {#2 #3} {\bf #4}, #5 (#6)}}}
\newcommand{\doie}[4]{\href{https://doi.org/#1}{{\cob {\bf #2}, #3 (#4)}}}
\newcommand{\doije}[4]{\href{https://doi.org/#1}{{\cob #2, #3 (#4)}}}
\newcommand{\doinn}[5]{\href{https://doi.org/#1}{{\cob {#2} {\bf #3}, #4 (#5)}}}
\newcommand{\doij}[5]{\href{https://doi.org/#1}{{\cob {#2} {#3} (#5) #4}}}

\newcommand{\ndoinn}[5]{\href{#1}{{\cob {#2} {\bf #3}, #4 (#5)}}}

\newcommand{\edbooks}[5]{\emph{#2}, edited by #1 (#3, #4, #5)}
\newcommand{\procsinm}[5]{in \emph{#1}, edited by #2 (#3, #4, #5)}

\newcommand{\tia}[1]{{#1},}

\renewcommand{\leq}{\leqslant}
\renewcommand{\geq}{\geqslant}
\newcommand{\Eq}[1]{(\ref{#1})}
\newcommand{\Eqq}[1]{Eq.~(\ref{#1})}

\def\rme{e}
\def\rmd{d}

\def\a{\alpha}
\def\b{\beta}
\def\de{\delta}

\def\g{\gamma}
\def\la{\lambda}

\def\e{\epsilon}

\def\om{\omega}
\def\G{\Gamma}
\def\t{\tau}
\def\s{\sigma}

\def\N{\nabla}
\def\B{\Box}

\def\mst{M_*}

\def\lst{\ell_*}

\def\cA{\mathcal{A}}

\def\cF{\mathcal{F}}

\def\cH{\mathcal{H}}

\def\cL{\mathcal{L}}

\def\cP{\mathcal{P}}

\def\cR{\mathcal{R}}
\def\cS{\mathcal{S}}

\def\cob{\color{blue}}

\begin{document}
	
	\title{Hawking area law in quantum gravity}
	
	\author{Gianluca Calcagni\,\orcidlink{0000-0003-2631-4588}}
	\email{g.calcagni@csic.es}
	\affiliation{Instituto de Estructura de la Materia, CSIC, Serrano 121, 28006 Madrid, Spain}
	
\begin{abstract}
We show that the LIGO--Virgo--KAGRA verification of Hawking area law carries profound consequences for quantum gravity if such a law is postulated to hold exactly. The observed mergers can be produced in local Stelle gravity and in nonlocal quantum-gravity theories with entire or fractional form factors either by (i) singular Ricci-flat black holes or (ii) possibly regular classical black holes under very restrictive conditions: absence of $R^2$ and (Riemann)${}^2$ terms in the action, absence of extra real poles in the graviton propagator, and positivity of its spectral representation. To date, this is the strongest simplification of the ambiguities of this class of theories. We also prove that the classical standard entropy-area law holds as a consequence of Hawking area law and provide a rigorous realization of Barrow's fractal black holes otherwise.
\end{abstract}

\preprint{\doin{10.1103/xm1k-65zs}{PHYSICAL REVIEW}{D}{114}{L021906}{2026} [\arX{2604.18669}]\hspace{6.5cm} April 19, 2026}
	
\maketitle


While intuition easily admits that observations related to black holes can effectively limit long-range modifications of Einstein gravity, it is far less obvious that they can also put quantum gravity \cite{Bam24} to test. The radius of an intermediate-mass black hole of $100\,M_\odot$ solar masses is about 40 orders of magnitude larger than the Planck length and one might naively think that gravitational-wave (GW) observations would be insensitive to any microscopic alteration of Einstein theory. This is not true thanks to the astounding precision reached in the measurement of mass (mirror) displacements in interferometers when a GW passes through. Indeed, GW astronomy has long since been recognized as a unique window into the behavior of classical or even quantum gravity \cite{Damour:1993id,Will:1994fb,Ryan:1997hg,Will:1997bb,Damour:1998jk,Amelino-Camelia:1998mjq,Ng:1999hm,Scharre:2001hn,Dreyer:2003bv,Collins:2004ex,Will:2004xi,Hughes:2004vw,Berti:2005qd,Berti:2005ys,Alexander:2007kv,Schutz:2009tz,Nishizawa:2009bf,Arun:2009pq,Sopuerta:2009iy,Stavridis:2009mb,Yagi:2009zm,Cannella:2009he,Yunes:2009ke,Nishizawa:2009jh,Cornish:2011ys,Mirshekari:2011yq,Yagi:2011xp,Yunes:2011aa,Chatziioannou:2012rf,Canizares:2012is,Yagi:2012ya,Yagi:2012vf,Mirshekari:2013vb,Nishizawa:2014zra,Berti:2015itd,Isi:2015cva}, and, after the first discovery of a merger event by LIGO--Virgo \cite{Abbott:2016blz} and the posterior detection of a GW background by the International Pulsar Timing Array \cite{NANOGrav:2023gor,NANOGrav:2023hvm,EPTA:2023fyk,Reardon:2023gzh,Xu:2023wog,InternationalPulsarTimingArray:2023mzf,Yu:2025dor}, this expectation has been met and constantly updated with current and projected constraints \cite{Ellis:2016rrr,Kostelecky:2016kfm,Calcagni:2016zqv,Arzano:2016twc,Yunes:2016jcc,Gasperini:2016gre,Kobakhidze:2016cqh,Amelino-Camelia:2017pdr,Nishizawa:2017nef,Ezquiaga:2017ekz,Cabero:2017avf,Wang:2017igw,Belgacem:2017ihm,Berti:2018cxi,Bosso:2018ckz,Tahura:2018zuq,Nishizawa:2018srh,Addazi:2018uhd,Maselli:2018fay,Berry:2019wgg,Calcagni:2019kzo,Hagihara:2019ihn,Giddings:2019ujs,Nair:2019iur,LISACosmologyWorkingGroup:2019mwx,Calcagni:2019ngc,Carson:2019kkh,Agullo:2020hxe,Calcagni:2020tvw,Takeda:2021hgo,LISACosmologyWorkingGroup:2022wjo,Calcagni:2022tuz,Takeda:2023wqn,Ben-Dayan:2024aec,Loutrel:2025bqn,Curras:2025hmo,Mu:2026pxg}. LIGO--Virgo--KAGRA (LVK) data have been accumulating, up to the current catalog of more than 300 mergers \cite{LIGOScientific:2026sit,LIGOScientific:2026ctl}, and made it possible to place stringent bounds on anything deviating from Einstein gravity, from the mass and speed of the graviton to violations of Lorentz invariance \cite{TheLIGOScientific:2016src,LIGOScientific:2019fpa,LIGOScientific:2026fcf,LIGOScientific:2026uyd}. So far, there is no evidence of new physics but the field is in rapid evolution and we have not exhausted all its possibilities. Third-generation detectors such as the Laser Interferometer Space Antenna \cite{Colpi:2024xhw} and Einstein Telescope \cite{Abac:2025saz} are further expanding the battery of tests and the range of models of interest.

Among the aspects touched by GW observations with momentous ramifications both for general relativity and for quantum gravity is Hawking area law for classical black holes \cite{Hawking:1971tu}. This law (reviewed below) states that the area of the event horizon of a classical black hole cannot decrease in time. GWs have been argued to be capable of testing this law \cite{Hughes:2004vw,Cabero:2017avf}, and this was accomplished in 2025 when the LVK Collaboration reported the data of event GW250114 \cite{LIGOScientific:2025rid}. Assuming a Kerr geometry, waveform models indicate this to be the merger of two black holes of masses $m_1= 33.6^{+1.2}_{-0.8} M_\odot$ and $m_2= 32.2^{+0.8}_{-1.3} M_\odot$ and spins $\chi_1\leq 0.24$  and $\chi_2\leq 0.26$, that coalesced into a final black-hole configuration with $m_{\rm f}= 62.7^{+1.0}_{-1.1} M_\odot$ and $\chi_{\rm f}=0.68^{+0.01}_{-0.01}$. From the masses and the spins, they calculated the event-horizon areas $\cA_1$, $\cA_2$, and $\cA_{\rm f}$ using Kerr's area formula and showed that
\be\label{area}
\cA_1+\cA_2 < \cA_{\rm f} \qquad \textrm{at $3.4\s$ level}\,,
\ee
about $88.5\,{\rm ms}$ before the peak of the averaged GW strain, with credibility level increasing above $5\s$ about $3.5\,{\rm ms}$ before the peak. These results greatly improve 
previous constraints \cite{Isi:2020tac,Correia:2023ipz,Tang:2025jyj} 
and, by extrapolation, they are taken as evidence that the area $\cA$ of a classical black-hole event horizon grows with time, as predicted by the area law. 

These data can be used to constrain violations of the area law, which are expected at the semiclassical level by evaporation via Hawking radiation \cite{Hawking:1975vcx,Page:2004xp}, but here we make a more minimalistic observation: If taken as an exact law, the growth of the area of the event horizon is sufficient to place remarkable constraints on quantum gravity with higher-curvature terms, in particular, a large class of field theories with nonlocal operators. 
We consider a generic nonlocal gravitational action
\ba
S\!&=&\!\frac{\Mpl^2}{2}\!\int\!\rmd^4x\,\sqrt{|g|}\big[R+R\cF_R(\B) R+G_{\mu\nu}\cF_2(\B) R^{\mu\nu}\nn
&&\qquad\qquad\qquad\qquad\!\!\!\!+R_{\mu\nu\s\t}\cF_4(\B) R^{\mu\nu\s\t}+\cL_{\rm K}\big],\label{nqg}
\ea 
where $\Mpl^2=(8\pi G)^{-1}$ is the reduced Planck mass, $G_{\mu\nu}=R_{\mu\nu}-(1/2) g_{\mu\nu}R$ is Einstein's tensor, $\cF_{R,2,4}(\B)$ are either constant (Stelle gravity \cite{Stelle:1976gc,Stelle:1977ry}) or nonlocal form factors, and the so-called killer Lagrangian $\cL_{\rm K}=O(\cR^4)$ will be discussed at the end. In this class of theories, two main scenarios have received attention: nonlocal quantum gravity (NLQG) made of entire, in particular asymptotically polynomial, form factors \cite{Kuzmin:1989sp,Modesto:2011kw,Biswas:2011ar,Biswas:2013cha,Modesto:2014lga,Modesto:2015lna,Dona:2015tra,Modesto:2016max,Calcagni:2018gke,Briscese:2018oyx,Buoninfante:2018mre,Modesto:2022asj,Calcagni:2022tuz,Calcagni:2023goc,Calcagni:2024xku,Briscese:2024tvc} (see \cite{Modesto:2017sdr,Buoninfante:2022ild,BasiBeneito:2022wux,Koshelev:2023elc,CaMo} for reviews) and fractional quantum gravity (FQG) made of fractional operators \cite{Calcagni:2024xku,Calcagni:2021ljs,Calcagni:2021ipd,Calcagni:2021aap,Calcagni:2022shb,Calcagni:2025wnn,Briscese:2026jcf,Calcagni:2026sfx}. Take $\cF_2$ as a reference (the other form factors $\cF_{R,4}$ are equal or proportional to $\cF_2$).
In NLQG,
\be\label{FNLQG}
\textrm{NLQG:}\qquad \cF_2(\B)=\frac{\rme^{{\rm H}_2(\B)}-1}{\B}\,,
\ee
where ${\rm H}_2(z)={\rm Ein}(z)=\ln p(z)+\G[0,p(z)]+\g_{\textsc{e}}$ is called complementary exponential integral, $\g_{\textsc{e}}$ is the Euler--Mascheroni constant, $p$ is a polynomial of $\lst^2\B$ of degree $n_p$, and $\lst\equiv \mst^{-1}$ is a fundamental length scale. It is not difficult to show that the graviton kinetic term in the linearized equations of motion becomes $\B\rme^{{\rm H}_2(\B)} h_{\mu\nu}$. At high energies, $\rme^{{\rm H}_2(\B)}\simeq \rme^{\g_{\textsc{e}}}p(\B)$. All that follows would also hold for other entire form factors such as Wataghin (${\rm H}_2=-\lst^2\B$) \cite{Wataghin:1934ann} and Krasnikov (${\rm H}_2=\lst^4\B^2$) \cite{Krasnikov:1987yj}, but these cannot be used in quantum gravity due to a convergence issue of Feynman diagrams typical of theories with gauge symmetries \cite{BasiBeneito:2022wux,CaMo}.

In FQG, nonlocal operators form a universality class whose elements have the same scaling in asymptotic regimes \cite{Briscese:2026jcf,Calcagni:2026sfx}. For instance, the so-called Hermitian simple form factor
\be\label{Ffqg}
\textrm{FQG:}\qquad \cF_2(\B)=\lst^2(\lst^4\B^2)^{\frac{\g}{2}-1}\,,
\ee
gives the prototypical ``$\B+\B^\g$'' graviton kinetic term $[\B+\lst^{-2}(\lst^4\B^2)^{\g/2}]h_{\mu\nu}$. 

In classical Einstein gravity, the FHawking area theorem can be sketched as follows. Consider a $(D-1)$-dimensional hypersurface $\cH$ and a congruence of null geodesics normal to it with tangent $k^\mu$ ($k_\mu k^\mu=0$) and affine parameter $\la$. In particular, assuming cosmic censorship (no naked singularities), we can identify $\cH$ with the event horizon of a black hole.
Decompose the metric $g_{\mu\nu}=h_{\mu\nu}-2n_{(\mu}k_{\nu)}$, where $n_\mu$ is an auxiliary null vector ($n_\mu n^\mu=0$) such that $n_\mu k^\mu=-1$. Define as usual the covariant expansion $\theta\coloneqq \N_\mu k^\mu$, the shear $\s_{\mu\nu}\coloneq h_\mu^\s h_\nu^\t \N_{(\s} k_{\t)}-\theta h_{\mu\nu}/2$, and the vorticity $\om_{\mu\nu}\coloneqq\N_{[\nu} k_{\mu]}$ ($=0$ on $\cH$). The Raychaudhuri equation for the null congruence $k^\mu$ is 
\be\label{ray}
\frac{\rmd\theta}{\rmd\la}=-\frac{\theta^2}{D-2}-\s_{\mu\nu}\s^{\mu\nu}+\om_{\mu\nu}\om^{\mu\nu}-R_{\mu\nu} k^\mu k^\nu\,.
\ee
This expression is purely kinematical but it can be further manipulated by including information on the dynamics. From the Einstein equations, $R_{\mu\nu} k^\mu k^\nu=\Mpl^2T_{\mu\nu} k^\mu k^\nu$, where $T_{\mu\nu}$ is the energy-momentum tensor. Assuming the null energy condition (NEC) $T_{\mu\nu} k^\mu k^\nu\geq 0$ and noting that $\s_{\mu\nu}\s^{\mu\nu}\geq 0$, for $D=4$ we get
\be
\frac{\rmd\theta}{\rmd\la}\leq -\frac12\theta^2\quad \Longrightarrow\quad \frac{1}{\theta(\la)}\geq \frac{1}{\theta(\la_0)}+\frac12(\la-\la_0)\,,
\ee
where we integrated from some $\la_0$. As a consequence, if $\theta(\la_0)<0$, then $\theta(\la)\to -\infty$ at finite $\la$ and the congruence develops a caustic at a finite affine parameter. However, the event horizon is the boundary of the causal past of future null infinity and is therefore generated by null geodesics that have no future end points. Therefore, it must be $\theta(\la)\geq 0$ everywhere on $\cH$. On the other hand, the expansion can also be expressed as $\theta=\rmd\ln\cA/\rmd\la$, where $\cA(\la)$ is the area of a spacelike slice of $\cH$. Since $\theta\geq 0$, Hawking area theorem is proven:
\be\label{7}
\frac{\rmd\cA}{\rmd\la} =\theta\cA\geq 0\,.
\ee
Hence, the area of the event horizon of a classical black hole cannot decrease with the affine parameter, or, more colloquially, it increases in time.

The theorem does not assume any specific background but, in exchange, it uses the Einstein equations and the NEC. Trading generality of the background for generality of the dynamics leads to an immediate application of the theorem also to any theory \Eq{nqg} with $\cF_4=0$. In the absence of the (Riemann)${}^2$ term, Ricci-flat ($R_{\mu\nu}=0$) black holes such as Schwarzschild and Kerr are exact solutions of the equations of motion of \Eq{nqg}. These are known solutions of Stelle gravity, NLQG and FQG \cite{Stelle:1977ry,Briscese:2019rii,Briscese:2026jcf} when the $\cF_4$ term is dropped. (Then, the problem of singularity resolution is solved at the quantum level by considering finite theories with exact Weyl invariance \cite{Modesto:2022asj,Bambi:2016wdn}.) As a consequence, these theories are compatible with the LVK observations and, if gravity was described by a field theory of the above, then the observed mergers would be generated by singular black holes.

When $\cF_4\neq 0$, only approximate (linearized) black-hole solutions have been explored and they have been found to be regular, thus pointing towards a resolution of the singularity problem already at the classical level \cite{Stelle:1977ry,Cornell:2017irh,Buoninfante:2018xiw,Koshelev:2018hpt,Buoninfante:2018stt,Buoninfante:2018rlq}. However, no such background can sustain the area theorem because it does not guarantee that $R_{\mu\nu} k^\mu k^\nu\geq 0$. Indeed, adding matter to \Eqq{nqg}, the generalized Einstein's equations in these theories are \cite{Biswas:2013cha,Calcagni:2021aap,Zhao:2023tox}
\be\label{Emn}
\cP(\B)G_{\mu\nu}+ B_{\mu\nu}=\Mpl^2 T_{\mu\nu}\,,
\ee
where $\cP(\B)\coloneqq 1+\cF_2(\B)\B$ and $B_{\mu\nu}$ is a complicated nonlocal operator with $O(\cR)$ and $O(\cR^2)$ derivative terms:
\ba
\hspace{-.35cm}B_{\mu\nu} \!\!\!&=&\!\! 2[g_{\mu\nu}\B-\N_{(\mu}\N_{\nu)}]\cF_RR+g_{\mu\nu}\N^\s\N^\t\cF_2 G_{\s\t}\nn
\hspace{-.35cm}&&\!\!-2\N^\s\N_{(\mu}\cF_2 G_{\nu)\s}-8\N^\s\N^\t\cF_4 R_{\mu\s\t\nu}+O(\cR^2).\label{Bmn}
\ea
Since, in general, $B_{\mu\nu} k^\mu k^\nu$ has no definite sign, focusing of geodesics is not granted. However, we can extract a sufficient condition on the form factors in order to satisfy the area theorem. To this aim, we drop $O(\cR^2)$ terms in $B_{\mu\nu}$ and consider a low-curvature regime, which is adequate for stellar-origin or supermassive black holes. (The Kretschmann invariant at the Schwarzschild radius $r_{\rm s}=2Gm$ is $R_{\mu\nu\s\t}R^{\mu\nu\s\t}\propto r_{\rm s}^2r^{-6}= r_{\rm s}^{-4}$; this is large only for microscopic objects.) Contracting \Eq{Emn} and \Eq{Bmn} with the null bivector $k^\mu k^\nu$ and using the contracted Bianchi identities $\N_\s R^\s_\nu=\N_\nu R/2$, we get
\ban
\Mpl^2 T_{\mu\nu}k^\mu k^\nu 
&=&\cP(\B) R_{\mu\nu}k^\mu k^\nu-2(k^\mu\N_\mu)^2\cF_RR\nn
&& -8\N^\s\N^\t\cF_4 R_{\mu\s\t\nu} k^\mu k^\nu+O(\cR^2)\,,
\ean
where we absorbed the curvature piece of the commutator $[\N_\s,\N_\mu] X^\s=-R_{\mu\t}X^\t$ into the $O(\cR^2)$ term. The latter introduces higher powers of $k\cdot\N$ and additional sign-indefinite pieces that can only worsen the situation unless highly constrained or, as under our assumptions, subdominant. Thus, the sign of $R_{\mu\nu} k^\mu k^\nu$ is governed by
\ba
R_{\mu\nu}k^\mu k^\nu &\simeq& \frac{1}{\cP(\B)} \Mpl^2 T_{\mu\nu}k^\mu k^\nu-8\frac{\cF_4}{\cP(\B)}k^\mu k^\nu\B R_{\mu\nu} \nn
&& + 2\frac{(\cF_R+2\cF_4)(k\cdot\N)^2}{\cP(\B)}R \,,\label{final}
\ea
where we dropped $O(\cR^2)$ terms and it is necessary that $\cP(\B)$ be invertible, i.e., $\cP(\la)\neq 0$ for any $\la$ in the spectrum of $\B$. This is clearly true in NLQG, where
\be\label{PNLQG}
\textrm{NLQG:}\qquad \cP(\B)=\rme^{{\rm H}_2}
\ee
is an entire function. In Hermitian simple FQG, we have
\be\label{Pfqg}
\textrm{FQG:}\qquad \cP(\B)=1+(\lst^2\B)^{-1}(\lst^4\B^2)^{\frac{\g}{2}}\,.
\ee
The complex roots of $\cP(\la)=0$ do not affect the focusing condition directly because classical solutions are built from real modes and the operator is evaluated on a real spectral support. Therefore, invertibility of \Eq{Pfqg} holds for all real $\la$ in the spectrum of $\B$ provided $\g>1$ (no divergence at the zero mode $\la=0$) and has no new zeros, i.e., no new real poles in the Green's function $[\cP(\B)\B]^{-1}$. This is the case only when the Riemann sheet of FQG is chosen to avoid real poles, which correspond either to a tachyon or to a heavy massive mode with mass $\approx \mst=\lst^{-1}$ \cite{Calcagni:2026sfx}. Here we are acquiring an empirical justification to the removal of the heavy mode from the spectrum. While this mode can be allowed in particle-physics phenomenology since it would only appear at very high (possibly grand-unification or Planckian) energies $\mst$, the survival of Hawking area law in FQG is tied to the absence of such degree of freedom.






The $\cF_R$ and $\cF_4$ terms in \Eqq{final} vanish for Ricci-flat solutions. On a general background, they are present but negligible for solar-mass or heavier black holes. An estimate of the size of these terms for a nonflat deformation of Schwarzschild with curvature $\cR\sim r_{\rm s}r^{-3}$ near the horizon is the following. At leading order in the fundamental scale $\mst=10^{-x}\Mpl$, $\cF_{R,4}= c\mst^{-2}+O(\N^2)$ for NLQG and $\cF_{R,4}\sim \mst^{-2(\g-1)}\N^{2(\g-2)}$ for FQG, where $c=O(1)$, so that for $m=10M_\odot=10^{39}\Mpl$ we have $\cF_{R,4} \N^2\cR\sim 10^{2x}(\Mpl/m)^4\Mpl^2\approx 10^{2x-152}\Mpl^2$ for NLQG and $\sim 10^{2(\g-1)x}(\Mpl/m)^{2\g}\Mpl^2\approx 10^{2(\g-1)x-76\g}\Mpl^2$ for FQG, unobservable even in the most optimistic case $x=15$ ($\mst=10\,{\rm TeV}$). However, if we conjecture the exact validity of Hawking area law, then the observed physics \Eq{area} of black-hole mergers is suggesting from \Eq{7} that it is sufficient to take 
\be\label{F04}
\cF_R=0\,,\qquad \cF_4=0\,,
\ee
if these macroscopic black holes are described by non-Ricci-flat solutions in a regime of small curvature. This result is of no-go type: miraculous cancellations could happen in \Eqq{final} and conspire to give \Eq{7}, but only for special solutions. Note that, both in NLQG and FQG, $\cF_R$ and $\cF_4$ can be set identically to zero without affecting unitarity and renormalizability \cite{Modesto:2011kw,Modesto:2014lga,Calcagni:2021aap,Calcagni:2022shb} ($\cL\propto R+G_{\mu\nu}\cF_2 R^{\mu\nu}$ Lagrangians were originally proposed in \cite{Modesto:2011kw,Barvinsky:2003kg,Barvinsky:2005db,Khoury:2006fg,Barvinsky:2011hd,Modesto:2013ioa}). 

Finally, the first term in \Eq{final} must maintain the sign of the NEC, that is, the inverse nonlocal operator must preserve positivity of $T_{\mu\nu}k^\mu k^\nu$. This implies that the operator $\cP^{-1}(\B)$ should have a positive spectral density. Again, this is trivial in NLQG, where the inverse of \Eq{PNLQG} is entire and positive. The FQG case is more complicated and we do not solve it here. The inverse of \Eq{Pfqg} is proportional to the Green's function $[\cP(\B)\B]^{-1}$, which was calculated in \cite{Briscese:2026jcf,Calcagni:2026sfx}. Removing the pole in zero from that expression, one gets the spectral representation of $\cP^{-1}(\B)$,
\ben
\frac{1}{\cP(\B)}\sim 1+\B\int_0^{+\infty}\rmd s\,\frac{\rho(s)}{s^2+(\lst^2\B)^2}\,,
\een
which should be applied to \Eqq{final}. The right-hand side is the superposition of complex-conjugate pairs with a certain spectral density $\rho(s)$ \cite{Calcagni:2026sfx}. Such continuum of modes is made purely virtual at all orders in perturbation theory via the Anselmi--Piva prescription \cite{Anselmi:2017yux,Anselmi:2017lia,Anselmi:2018bra,Anselmi:2019rxg,Anselmi:2021hab,Anselmi:2022toe,Anselmi:2025uzj,Anselmi:2025uda} and the classical dynamics is obtained by integrating out the virtual modes from the \emph{interim} action \Eq{nqg} \cite{Anselmi:2018bra,Anselmi:2019rxg,Anselmi:2025uda}. The sign of $\rho(s)$ in the \emph{interim} system does not directly connect with the positivity condition in the classicized one, and the calculation of the classicized dynamics of FQG is left for the future.


These conclusions do not rely too strongly on the NEC. The area theorem can be generalized and the NEC relaxed to weaker conditions allowing for a certain amount of negative energies, such as an averaged NEC \cite{Lesourd:2017hzs} or a negative, finite lower bound on the energy \cite{Kontou:2023ntd}. The latter places a bound on the evaporation rate (hence on the area decrease rate) of semiclassical black holes \cite{Kontou:2023ntd}. Weakening the sign bound on the energy does not help nonlocal theories much because the problematic terms in \Eqq{final} can in principle cross any such finite bound by an arbitrarily larger contribution.

So far, we have expressed all our results in terms of the event-horizon area $\cA$, with no reference to the thermodynamics of black holes. We now extend the definition of the Wald entropy to nonlocal theories; this was done already for spherically symmetric black holes \cite{Conroy:2015wfa,Myung:2017axf} but here we derive the result in a fully covariant way, dropping the assumption of spherical symmetry, for form factors not necessarily entire, and with a different physical interpretation. Wald's definition \cite{Wald:1993nt,Jacobson:1993vj,Iyer:1994ys,Iyer:1995kg}
\be
\cS = -2\pi\int_{\cH} \frac{\de \cL}{\de R_{\mu\nu\s\t}} \e_{\mu\nu}\e_{\s\t}\,,
\ee
where $\e_{\mu\nu}=k_\mu n_\nu-k_\nu n_\mu$ is the binormal to the horizon bifurcation surface $\cH$, was originally applied to local theories but it can also be used for covariant nonlocal theories such as those considered here. Thanks to the fact that the binormal $\e_{\mu\nu}$ is covariantly constant on an horizon cross-section $\Sigma\subset\cH$, notwithstanding the presence of the form factors $\cF_i$ one can use the identities $(\de R/\de R_{\mu\nu\s\t})\e_{\mu\nu}\e_{\s\t}=g^{\mu[\s}g^{\t]\nu}\e_{\mu\nu}\e_{\s\t}=\e_{\mu\nu}\e^{\mu\nu}=-2$, $(\de R_{\a\b}/\de R_{\mu\nu\s\t})\e_{\mu\nu}\e_{\s\t}=g^{\mu[\s}\de^{\t]}_\a\de^\nu_\b\e_{\mu\nu}\e_{\s\t}=2 k_{(\a} n_{\b)}$, $2R^{\a\b}k_\a n_\b=R_\Sigma-R$, and $R^{\mu\nu\s\t}\e_{\mu\nu}\e_{\s\t}=4R^{\mu\nu\s\t}k_\mu n_\nu k_\s n_\t=4R_\Sigma-2R$, where $R_{\Sigma}$ is the intrinsic Ricci curvature of any stationary Killing horizon. In four dimensions, we get
\ba
\cS&=&\frac{\cA}{4G}+\frac{1}{4G}\int_\Sigma \rmd^2x\,\sqrt{h}\,2(\cF_R+\cF_4) R\nn
&&-\frac{1}{4G}\int_\Sigma \rmd^2x\,\sqrt{h}\,(\cF_2+4\cF_4) R_{\Sigma}\,.\label{wald}
\ea
For a spherically symmetric horizon, $R_{\Sigma}={\rm const}$ and the last line in \Eqq{wald} is a constant shift that can be reabsorbed in the definition of $\cS$ and, from now on, ignored irrespective of the details of the form factors $\cF_{2,4}$. [In FQG, a stronger result holds since constants lie in the kernel of fractional form factors and this $O(R_{\Sigma})$ term vanishes identically]. The term $\cA/(4G)$ in \Eqq{wald} is the one of Einstein gravity. For constant $\cF_i\neq 0$, we recover the entropy-area law for Stelle gravity \cite{Jacobson:1993vj}, a special case of Lovelock gravity \cite{Jacobson:1993xs}. In particular, for $\cF_R={\rm const}$ and $\cF_{2,4}=0$, we reobtain the entropy of Starobinsky gravity \cite{Briscese:2007cd,Dyer:2008hb}. For nontrivial operators $\cF_i\neq 0$, we recover the result of \cite{Conroy:2015wfa} [their form factors $F_i$ are related to ours by $\cF_R=F_1+F_2/2$, $\cF_{2,4}=F_{2,3}$, so that $(2\cF_R+2\cF_4)R=(2F_1+F_2+2F_3)R$]. On Ricci-flat backgrounds or when \Eqq{F04} holds, 
 \Eqq{wald} matches the Bekenstein--Hawking entropy-area law $\cS=\cA/(4G)$ in Einstein gravity \cite{Bekenstein:1973ur,Hawking:1974rv}. This simplification was obtained in \cite{Conroy:2015wfa} by forbidding extra degrees of freedom apart from the graviton in the presence of entire form factors but, here, it rather comes from the removal of the $\cF_R$ and $\cF_4$ terms, which not only are unnecessary in the quantum theory but also violate Hawking area law. From our perspective, the obtention of exactly the standard Wald entropy in nonlocal theories (including fractional ones, which do not have entire $\cF_i$) is a consequence of elevating the empirically verified increase of the area in time  to an exact law. If the area increases, so does the Wald entropy. If $\cF_R\neq 0\neq\cF_4$, then neither the area nor the entropy grow monotonically.

Let us now examine other consequences of giving up Hawking area law as exact at all scales ($\cF_R\neq 0\neq \cF_4$). We show that the entropy \Eq{wald} has a resemblance to Barrow entropy \cite{Barrow:2020tzx} but also that they are essentially different. Barrow entropy is $\cS\sim \cA^{\dh^{\rm s}/2}$, where $2\leq\dh^{\rm s}\leq3$ is the Hausdorff dimension of the horizon surface of a fractal black hole embedded in a four-dimensional smooth spacetime. In contrast, in FQG the whole spacetime is a multifractal in a precise sense. The Hausdorff dimension of spacetime is $\dh=4$ at all scales, the spectral dimension is $\ds\simeq 4/\g<2$ in the ultraviolet \cite{Briscese:2026jcf,Calcagni:2012rm}, and the walk dimension is $\dw\simeq 2\g$ \cite{Calcagni:2012rm}. Therefore, spacetime is a weak multifractal \cite{Calcagni:2016azd} because it is almost everywhere differentiable and obeys the condition $\dw=2\dh/\ds$, which is part of the definition of fractal sets \cite{Calcagni:2016edi}. Black holes are not rough \emph{\`a la} Barrow and their event horizon has $\dh^{\rm s}=2$ because the observer measures the geometry with rulers following the same multiscaling as the ambient space \cite{Calcagni:2016azd,Calcagni:2016edi}. On a quasi-Schwarzschild background, $\int\rmd^2x\,\sqrt{h} \,\cF_i R
\sim\! \int\rmd r\, r^{2-2\g}\sim r^{3-2\g}$ and 
\be\label{bar}
\cS\sim \cA+\cA^{\frac{3}{2}-\frac{4}{\ds}}\,.
\ee
All the above also applies to NLQG with the replacement $\g\to n_p+1$ ($n_p\geq 4$ is the degree of $p$). In all cases, the entropy and the area do not grow together at all scales. In particular, in Stelle and Starobinsky gravity $\cS\sim \cA+\cA^{-1/2}$.

To summarize, while Barrow entropy is a heuristic attempt to express the thermodynamics of a black hole whose geometry has become fuzzy due to generic fractal-like quantum-gravity effects \cite{tHooft:1993dmi,Carlip:2009kf,Carlip:2017eud,Carlip:2019onx}, the entropy \Eq{wald}--\Eq{bar} is the outcome of a top-down theory where fractal spacetime geometry is fully under control. Note that, in fact, known bounds on the entropy exponent $\dh^{\rm s}$ \cite{Xia:2024nmp} do not apply to $\ds$ since this term in \Eq{bar} is additive.

Curiously, \Eqq{bar} recovers Tsallis--Cirto nonadditive entropy $\cS\sim\cA^{\de}$ \cite{Tsallis:2012js} with $\de=3/2-4/\ds$. In particular,
the case $\de=3/2$ corresponds to the limit of an ultradiffusive spacetime ($\ds\to\infty$). Also, \Eqq{bar} radically differs from the entropy found in a Hamiltonian model with fractional spatial derivatives \cite{Jalalzadeh:2021gtq,Junior:2023fwb,Jalalzadeh:2025uuv}, where the Schwarzschild radius $r_{\rm s}(m)$ is no longer linear in $m$. We have not explored the physical meaning of all of this.

The result \Eq{wald} is purely classical. At the quantum level, since both NLQG and FQG are renormalizable and their divergences are local \cite{Modesto:2014lga,Modesto:2015lna,Calcagni:2022shb,Calcagni:2023goc}, no new terms are generated in \Eq{wald} by these divergences and the only effect is a renormalization of the coupling constants in $\cS$, with a mechanism identical to the one shown in \cite{Fursaev:1994ea}. Therefore, the entropy-area law is stable against perturbative quantum corrections. On the other hand, the nonfinite versions of these theories carry a conformal anomaly which, as is well-known in perturbative quantum gravity, produces a logarithmic correction $\cS\sim \cA +\ln \cA$ \cite{Fursaev:1994te,Cai:2009ua}. This correction is also common to approaches as diverse as string theory \cite{Carlip:2000nv,Banerjee:2011jp,Sen:2012cj} and supergravity \cite{Iliesiu:2022onk}, loop quantum gravity \cite{Kaul:2000kf}, asymptotic safety \cite{Falls:2012nd}, noncommutative spacetimes \cite{Gupta:2022oel}, and effective field theory \cite{Xiao:2021zly}. Although the $\ln\cA$ correction has all the characteristics of a universal feature in quantum gravity \cite{Carlip:2000nv}, it seems that NLQG and FQG might not partake of this universality in their finite versions, which are realized by augmenting the action with higher-curvature operators called killers because they force the beta functions to zero \cite{Modesto:2014lga,Modesto:2015lna,Calcagni:2022shb,Calcagni:2023goc}. The killer sector is of the form $\cL_{\rm K}=R^2 f_0(\B) R^2+R_{\mu\nu}R^{\mu\nu}f_2(\B)R_{\mu\nu}R^{\mu\nu}$, where $f_{0,2}(\B)\propto \B^{n_p-3}$ is local in NLQG and $f_{0,2}(\B)\sim \B^{\g-4}$ (when $\g>4$ and up to an infrared regulator) is nonlocal in FQG. Equation~\Eq{wald} is then augmented by the contributions
\ba
\cS_{\rm K}&=&\frac{1}{4G}\int_\Sigma \rmd^2x\,\sqrt{h}\,2R(2f_0 R^2+f_2R_{\mu\nu}R^{\mu\nu})\nn
&&-\frac{1}{4G}\int_\Sigma \rmd^2x\,\sqrt{h}\,(2f_2R_{\mu\nu}R^{\mu\nu}) R_{\Sigma}\,.\label{waldk}
\ea
The $O(R_\Sigma)$ term is reabsorbable as before. The first integral yields $\cS_{\rm K}\sim\cA^{\frac{1}{2}-\frac{4}{\ds}}$ and vanishes on Ricci-flat solutions or in the super-renormalizable, nonfinite versions of the theory ($f_0=0=f_2$), in which case one expects the logarithmic correction from the conformal anomaly. 

In conclusion, we have derived the entropy-area law for multifractal spacetimes and made the notion of Barrow entropy precise in Stelle and nonlocal quantum gravity. Motivated by LVK observations, if we take Hawking area law to be exact, then these scenarios are constrained to a Lagrangian in the tensor sector of the form $\cL\propto R+G_{\mu\nu}\cF_2 R^{\mu\nu}$, with $R^2$ and (Riemann)${}^2$ terms identically zero, 
 if black holes are described by non-Ricci-flat solutions. This is a drastic reduction of ambiguity in the definition of the dynamics of all these theories and demonstrates the constraining potential of GW observations in the realm of quantum gravity.

	

\medskip

\noindent \emph{Acknowledgments.} The author thanks S.~Kuroyanagi for crucially triggering the thoughts that led to this work and L.~Modesto for useful discussions. He is supported by Grant No.\ PID2023-149018NB-C41 funded by the Spanish Ministry of Science, Innovation and Universities MCIN/AEI/10.13039/501100011\-033. His work was made possible also through the support of the WOST, \href{https://withoutspacetime.org}{WithOut SpaceTime project}, supported by Grant ID 63683 from the John Templeton Foundation. The opinions expressed in this work are those of the author and do not necessarily reflect the views of the John Templeton Foundation.

\medskip

\noindent\emph{Data availability.} There are no publicly available
research data or software supporting this manuscript. Requests for further information or data should be sent to the authors.


\end{document}